\documentclass[apj]{emulateapj}
\usepackage{psfig}

\newcommand{\etal}{et al.}

\newcommand\asca{{\it ASCA\/}}

\newcommand\rosat{{\it ROSAT\/}}
\newcommand\chandra{{\it Chandra}}

\newcommand\xmm{{\it XMM-Newton}}

\def\nh{\hbox{$N_{\rm H}$}}

\def\snr{Kes~79}
\def\psr{\rm{PSR J1852$+$0040}}
\def\src{\rm{CXOU~J185238.6$+$004020}}

\def\simlt{\mathrel{\hbox{\rlap{\hbox{\lower4pt\hbox{$\sim$}}}\hbox{$<$}}}}
\def\simgt{\mathrel{\hbox{\rlap{\hbox{\lower4pt\hbox{$\sim$}}}\hbox{$>$}}}}

\submitted{Submitted February 14, 2005}
\accepted{March 17, 2005}

\shorttitle{Discovery of a 105~ms X-ray Pulsar in Kes~79}
\shortauthors{Gotthelf \etal}

\begin{document}

\title{Discovery of a 105~ms X-ray Pulsar in  Kesteven~79: \\
On the Nature of Compact Central Objects in Supernova Remnants}


\author{E. V. Gotthelf \& J. P. Halpern}
\affil{Columbia Astrophysics Laboratory, Columbia University, 
550 West 120th Street, New York, NY 10027-6601; eric@astro.columbia.edu,
jules@astro.columbia.edu}

\and 

\author{F. D. Seward}
\affil{Smithsonian Astrophysical Observatory, 60 Garden Street, 
Cambridge, MA 02138; fds@head.cfa.harvard.edu}

\begin{abstract}
We report the discovery of 105-ms X-ray pulsations from the compact
central object (CCO) in the supernova remnant \snr\ using data
acquired with the {\it Newton X-Ray Multi-Mirror Mission}.  Two
observations of the pulsar taken 6~days apart yield an upper limit
on its spin-down rate of $\dot P < 7 \times 10^{-14}$~s~s$^{-1}$ and
no evidence for binary orbital motion.  The implied energy loss
rate is $\dot E < 2 \times 10^{36}$~ergs~s$^{-1}$, surface magnetic
field strength is $B_{\rm p} < 3 \times 10^{12}$~G, and spin-down age
is $\tau > 24$~kyr.  The latter exceeds the remnant's estimated age,
suggesting that the pulsar was born spinning near its current
period. The X-ray spectrum of \psr\ is best characterized by a
blackbody model of temperature $kT_{BB} = 0.44\pm0.03$ keV, radius $R_{BB}
\approx 0.9$~km, and $L_{\rm bol} = 3.7 \times 10^{33}$ ergs~s$^{-1}$
at $d = 7.1$~kpc. The sinusoidal light curve is modulated with a pulsed
fraction of $>45\%$, suggestive of a small hot spot on the surface of
the rotating neutron star. The lack of a discernible pulsar wind
nebula is consistent with an interpretation of \psr\ as a
rotation-powered pulsar whose spin-down luminosity falls below the
empirical threshold for generating bright wind nebulae, $\dot E_{\rm
c} \approx 4 \times 10^{36}$~ergs~s$^{-1}$. The age discrepancy implies
that its $\dot E$ has always been below $\dot E_c$, perhaps a
distinguishing property of the CCOs. Alternatively, the X-ray spectrum
of \psr\ suggests a low-luminosity AXP, but the weak inferred $B_{\rm p}$
field is incompatible with a magnetar theory of its X-ray
luminosity. So far, we cannot exclude accretion from a fall-back disk.
The ordinary spin parameters discovered from \psr\
highlight the difficulty that existing theories of isolated neutron
stars have in explaining the high luminosities and temperatures 
of CCO thermal X-ray spectra. 
\end{abstract}

\keywords{stars: neutron  --- pulsars: (\src, \psr) --- supernova
remnants: individual (\snr) }

\section {Introduction}

The discovery in recent years of many isolated neutron stars (NSs) at
the centers of supernova remnants (SNRs) confirms the long-held notion
that these ultra-dense stellar remnants are born in supernova
explosions \citep{bad34}.  Most of these NSs are identified as
pulsars, whose emission derives either from rotational energy loss, as
for the rapidly spinning pulsars in the Crab ($P = 33$~ms) and Vela
($P = 89$~ms) remnants, or from magnetic field decay, as posited for
the slower ($5 < P \leq 12$~s) anomalous X-ray pulsars (AXPs) and
related objects (see Mereghetti 2000 for a review). However, the
nature of a small but growing collection of young ($\simlt 10^4$~yrs)
NSs in SNRs remains a mystery. These so-called Compact Central Objects
(CCOs) are seemingly isolated NSs, distinguished by their steady flux,
predominantly thermal emission, lack of optical or radio counterparts,
and absence of a surrounding pulsar wind nebula (see \citealt*{pav04}
for a review).

The six firm examples of CCOs are the central X-ray sources discovered
in Cas~A \citep{tan99}, Pup~A \citep*{pet82}, G347.3$-$0.5
\citep{sla99}, PKS~1209--51/52 \citep{hel84}, G266.2--1.2
\citep{sla01}, and Kes~79 \citep{sew03}. Their luminosities range over
$10^{33} \simlt L_{x} \simlt 10^{34}$ erg~s$^{-1}$, typical of the
younger pulsars; however their spectra are best characterized as hot
blackbody emission of $kT_{\rm BB} \sim 0.4$~keV, rather than by
power-law models. This is significantly hotter than radio pulsars or
other radio-quiet NSs \citep*[RQNSs; e.g.,][]{car96}, but similar to
the thermal component of the more luminous AXPs/SGRs. Distinct from
the rest, PSR~J1210--5226 in PKS~1209--51/52 is a $424$~ms pulsar
\citep{zav00} that displays large deviations in its spin-down rate, a
softer spectrum of $kT_{\rm BB} = 0.22$~keV, and cyclotron resonance
lines \citep*{san02,pav02,mer02,big03,zav04}. These properties
suggest that PSR~J1210--5226 is variously a wide-binary system, a strongly
glitching NS, or an accretor from fall-back material \citep{zav04}.

\nobreak \src\ is the recently discovered CCO in supernova remnant
Kes~79 (G33.6$+$0.1; \citealt{sew03}). The distance to this unresolved
\chandra\ source is estimated as $7.1$~kpc, derived from \ion{H}{1}
and OH absorption studies \citep{fra89,gre92} and updated using the
Galactic rotation curves of \citet{cas98}. Its X-ray luminosity all
but rules out any late-type star, and the spectral slope (photon index
$\Gamma = 4.2$; \citealt{sew03}) nominally eliminates an AGN
origin. The steady flux disfavors an accreting binary origin for the
X-ray emission, and the absence of a wind nebula argues against an
energetic rotation-powered pulsar.
In this paper we report the discovery of 105~ms pulsations from the
CCO associated with \snr, with an upper limit on the period derivative,
and we consider possible interpretations of its timing and spectral
properties.  We leave the analysis of the SNR to a subsequent paper.

\section{\xmm\ Observations}

The central source in \snr\ was observed twice with the {\it Newton
X-Ray Multi-Mirror Mission} (\xmm) on 2004 October 18 and 23.  We
analyze data from the European Photon Imaging Camera (EPIC;
\citealt{tur03}), which consists of three CCD imagers, EPIC~pn and two
EPIC~MOSs. These detectors are sensitive to X-rays in the nominal
$0.1-12$~keV range.  The EPIC~pn was operated in ``small window'' mode
with a $4\farcm3 \times 4\farcm3$ field-of-view (FOV) and 5.7~ms time
resolution that allows a search for even the most rapidly rotating
pulsar. The target was placed at the default EPIC~pn focal plane
location for a point source. The EPIC~MOS data were obtained in
``full frame'' mode with a $30^{\prime}$ diameter FOV to image the
diffuse supernova remnant emission and its environment. The time
resolution in this mode is 2.7~s. For all three cameras the medium
density filter was used.

We processed the data with Science Analysis System version SAS 6.0.0
(20040318\_1831), and screened the photon event lists using the
standard filter criteria.  Both observations were uncontaminated by
flare events and provided 30.5~ks of good EPIC~pn exposure time during
each epoch. After taking into account the CCD readout dead-time
(29\%), this translates to 21.6~ks of live-time per observation in
``small window'' mode. Each EPIC~MOS provided nearly 31~ks of good
exposure time per observation with no dead-time. In all cases, the
source count rates were too low for pile-up effects to be
significant. Photon arrival times were converted to the solar system
barycenter using the \chandra\ derived source coordinates, $18^{\rm
h}52^{\rm m}38\fs57, +00\arcdeg40\arcmin19\farcs8$ (J2000.0), which we
determined from the latest aspect analysis.

\begin{figure}[h]
\centerline{
\hfill
\psfig{figure=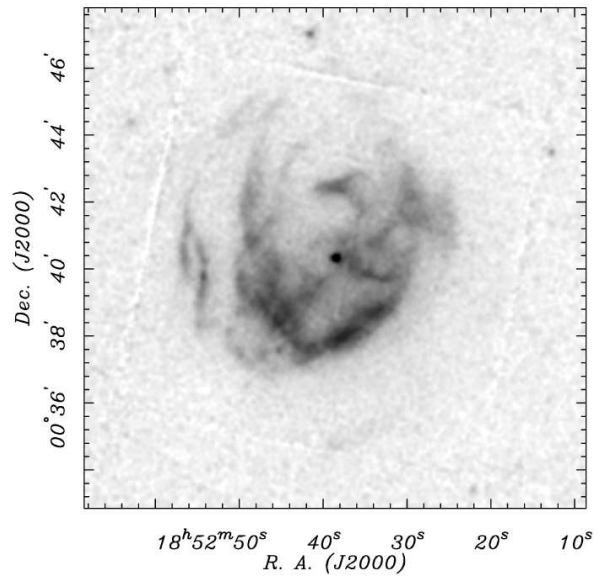,width=0.9\linewidth,angle=270}
\hfill
}
\caption{The 61.7~ks \xmm\ EPIC~MOS exposure-corrected
X-ray image ($0.3-10$ keV) of the shell-type supernova remnant Kes~79.
The image is smoothed with a 3-pixel boxcar algorithm and
scaled by the square-root of the intensity.  Gaps between the
CCDs are visible as linear artifacts.  Timing analysis of
the central point-source, \src, using the EPIC~pn data, reveals a
105~ms pulsar.}
\label{imageplot}
\end{figure}

Figure~\ref{imageplot} shows the exposure-corrected EPIC~MOS image of
\snr\ incorporating data from both \xmm\ observations, centered on
\src. The new data closely matches the earlier \chandra\ image
presented in Sun \etal\ (2004). In the vicinity of the pulsar the
surface brightness of the SNR varies greatly. This requires care in
choosing the source and background regions. In the following analysis
we choose a small source aperture $30^{\prime\prime}$ in diameter
centered on \src\ to minimize background contamination from the
SNR. This region encloses $\geq 80\%$ of the source flux. An annular
background region would be contaminated by the increased SNR emission
to the northwest of the pulsar. This is particularly troublesome for
the spectral analysis since the soft source spectrum overlaps that of
the thermal emission from the SNR. Instead, background counts were
accumulated from a relatively uniform, $30^{\prime\prime}$ diameter
region centered at coordinates $18^{\rm h}52^{\rm m}39\fs30,
+00\arcdeg39\arcmin49\farcs2$ (J2000.0), which is just to the
southeast of the source.


\section{Timing Analysis}

\begin{figure}[h]
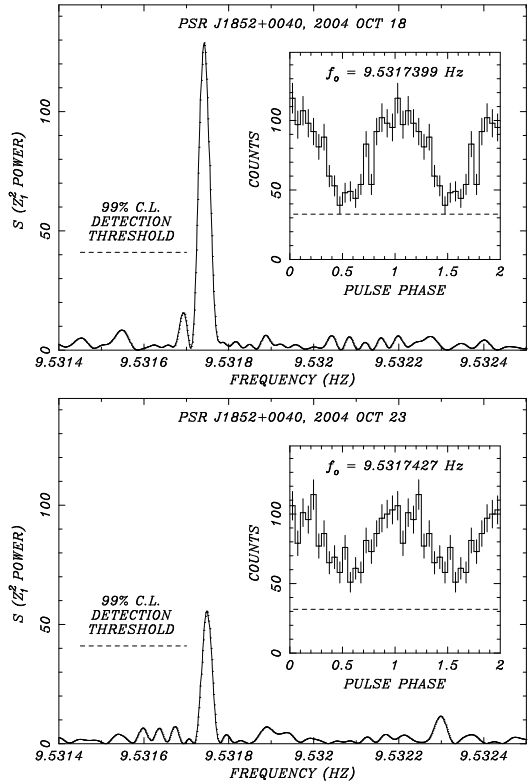

\centerline{
\hfill
\psfig{figure=f2a.eps,width=0.8\linewidth,angle=270}
\hfill
}
\vspace{0.1cm}
\centerline{
\hfill
\psfig{figure=f2b.eps,width=0.8\linewidth,angle=270}
\hfill
}
\caption{Discovery of \psr\ in \snr\ using \xmm\ EPIC~pn data acquired
on 2004 October 18 ({\it left}) and October 23 ({\it right}). A highly
significant signal is found in $1-5$ keV photons extracted from a
$30^{\prime\prime}$ diameter aperture at the location of \src. The detection
threshold for a blind search of a $2^{23}$-element FFT is
indicated. {\it Inset\/}: Folded light-curve of \psr. The background level
is indicated by the dashed line.}
\label{timeplot}
\end{figure}

Arrival times of photons in the energy range $0.3-10$~keV were extracted
from the source region and searched for a pulsed signal using the Fast
Fourier Transform (FFT) algorithm. A highly significant signal is
detected in the October~18 time series at $P = 105$~ms using a
$2^{23}$-bin transform. No higher harmonics were found. We constructed
a periodogram centered on this signal using the $Z^2_1$ (Rayleigh)
test \citep{buc83} and localized the pulsed emission to $P =
104.91264(4)$~ms with a peak statistic of $Z^2_1 = 130$, after refining
the energy range to an optimal $1-5$~keV energy band. 
The number in parenthesis is the 95\% confidence uncertainty
in the last digit computed using the Monte Carlo method described in
\citet*{got99}.  The detection significance corresponds to a negligible 
probability of chance occurrence.
The periodogram is shown in Figure~\ref{timeplot}
along with the roughly sinusoidal pulse profile folded at the best
period. A similar analysis of the 2004 October~23 time series produced
a notably weaker signal with $Z^2_1 = 56$ at $P = 104.91261(5)$~ms
(see Table~\ref{timetable}). 
In neither observation do we find evidence of
significant energy dependence in the pulse shape.


\begin{deluxetable}{lccccc}
\tablewidth{240pt}
\tablecaption{\xmm\ Timing Results for \psr}
\tablehead{
\colhead{Epoch}	     & \colhead{Expo} & \colhead{Bkg\tablenotemark{a}} & \colhead{Source\tablenotemark{b}} & \colhead{Period\tablenotemark{c}} & \colhead{$f_p$\tablenotemark{d}}  \\
\colhead{(MJD[TDB])} & \colhead{(s)}  & \colhead{(s$^{-1}$)}  & \colhead{(s$^{-1}$)} & \colhead{(s)} & \colhead{(\%)}
}
\startdata
 53296.00135994 & 30587 & 0.030(1) & 0.041(2) & 0.10491264(4) & 86(16)\\
 53301.98462573 & 30515 & 0.029(1) & 0.046(2) & 0.10491261(5) & 61(16) \\
\enddata

\tablenotetext{a}{\footnotesize Background rate obtained using a $r=0\farcm25$ aperture placed just southeast of the source region (see \S 2),
and corrected for dead-time.  Statistical ($\sqrt{N}$) uncertainty in the last 
digit is given in parentheses.}
\tablenotetext{b}{\footnotesize Background and dead-time corrected count rate for a $r=0\farcm25$ aperture.  Statistical ($\sqrt{N}$) uncertainty in the last digit is given in parentheses.}
\tablenotetext{c}{\footnotesize Period derived from a $Z^2_1$ test. Period uncertainty is 95\% confidence computed by the Monte Carlo method
described in \citet{got99}.}
\tablenotetext{d}{\footnotesize Pulsed fraction defined as $f_p \equiv N({\rm pulsed}) /  N({\rm total})$.}
\label{timetable}
\end{deluxetable}


The background-subtracted pulsed fraction is $86\pm 16\%$ and 
$61\pm 16\%$ for the two observations, respectively.
Here, we define the pulsed fraction as 
$f_p \equiv N({\rm pulsed})/N({\rm total})$,
where we choose the minimum of the folded light curves as
the unpulsed level.
The quoted uncertainties are derived by propagating the counting
statistics of the light curve, but the pulsed fraction is also
quite sensitive to systematic uncertainty in the background,
which contributes about half of the total counts.
The differences in $Z^2_1$ and $f_p$ between the two
observations are suggestive of a real change,
but not large enough to be considered reliable at this stage.
According to the prescription of
\citet*{pav99}, the two values of $Z^2_1$ are inconsistent
at approximately the $3\sigma$ level.  If there is real
variability of the pulse shape, it has important
implications for the origin of the emission.

The uncertainty on the period measurement from the two \xmm\
observations is too large to derive a significant spin-down rate. In
fact, the measured period {\it decreased} with time, although not
significantly, $\dot P = \Delta P / \Delta T = (-6 \pm 13) \times
10^{-14}$~s~s$^{-1}$.  The error range is the 95\% confidence interval
and is computed by propagating the uncertainties on the individual
period measurements.  We also searched the combined data set over a
plausible range of $P$ and $\dot P$, for increased sensitivity.  We
were not able to identify the correct period derivative unambiguously
from the multiplicity of possible solutions.  For an upper limit on
the period derivative of $\dot P < 7 \times 10^{-14}$~s~s$^{-1}$ we
can constrain the spin-down power of \psr\ to $\dot E < 2 \times
10^{36}$~ergs~s$^{-1}$, the inferred magnetic field to $B_{\rm p} < 3
\times 10^{12}$~G, and the characteristic age to $\tau \equiv P/2\dot
P > 24$~kyr, parameters that are typical of a rotation-powered pulsar
with a normal magnetic field strength.

The two observations of \psr\ also allow us to place an upper limit on its
radial acceleration over 6~days corresponding to $\Delta v_{\rm r} <
0.32$ km~s$^{-1}$.  Finally, we looked for acceleration on $1-8$~hr
timescales by cross-correlating the pulse profiles obtained from
sub-sections of the time series with a template profile constructed
from all the data.  For the first observation, we derive an upper
limit of $a_x \sin i < 0.01$ light-seconds, less than those found for
the AXPs \citep*{mer98}.  The data from the second observation are
somewhat less constraining.  These limits exclude a main-sequence
companion filling its Roche lobe for inclination angles $i >
10^{\circ}$, the latter having 98.5\% a priori probability.

Previous searches for pulsations from the direction of \snr\ were
inconclusive.
A recent observation at 1.4~GHz using the Parkes radio telescope
places an upper limit on a coherent signal of $L_{1400} \simlt
3$~mJy~kpc$^2$, comparable to many faint pulsars but 10 times higher
than that of the faintest detected (Camilo \etal, in preparation). We
also tested archival X-ray data, in particular a relatively long
(25~ks), continuous \rosat\ PSPC observation obtained on 1991
September 28. Given the background rates at the pulsar position, 
even if \psr\ were 100\% pulsed, the
effective pulsed fraction in the \rosat\ observation would be 
no greater than $f_p \approx 27\%$.
However, the data are not sensitive enough to reveal it.
Similarly, in two available \asca\ observations, one a dedicated 37~ks
observation of \snr, and the other a short (9~ks) Galactic ridge
survey field observation, the expected pulsar signal is masked by
SNR emission because of the $1^{\prime}$ mirror resolution.

\section{Spectral Analysis}

Source and background spectra from each data set were accumulated in
apertures described in \S 2.  For each observation, data from the two
EPIC MOS cameras (MOS1+MOS2) were combined and analyzed as a single
data set.  After verifying that there were no emission-line features
detectable, all spectra were grouped into bins containing a minimum of
40 counts (including background). EPIC~pn and EPIC~MOS spectra pairs
from each observation were fitted simultaneously using the XSPEC
package, with the normalization for each data free to vary
independently. This allowed for differences in the overall flux
calibration, apparent at the $\simlt 10\%$ level. Some instrumental
variation is possibly due to the effects of the distinct pixel size
and mirror point-spread function on measuring the diffuse SNR
emission over the source and background regions.

\begin{figure}[h]
\centerline{
\hfill
\psfig{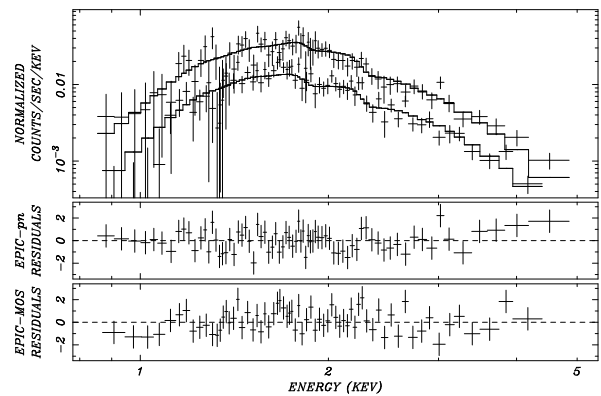}
\hfill
}
\caption{The \xmm\ spectrum of \psr\ in
Kes~79 from combined observations in 2004 October.
{\it Top panel\/}: Data from the EPIC~pn
({\it top curve}) and EPIC~MOS
({\it bottom curve}) cameras are shown with the best fitted blackbody model
using parameters given in Table~\ref{spectable}.
{\it Middle panel\/}: Residuals to the fit for the EPIC~pn.
{\it Bottom panel\/}: Residuals for the EPIC~MOS. The residuals are given in units of sigma.}
\label{specfig}
\end{figure}

The results of these fits using either an absorbed power-law or
blackbody model are presented in Table~\ref{spectable}. An acceptable
$\chi^2$ statistic is obtained using either model, for each epoch and
for the combined data. However, the blackbody model is preferred over
the power-law model based on its derived column density of $\nh =
(1.5\pm0.2) \times 10^{22}$~cm$^{-2}$, which is consistent with that
found for the remnant \citep[$\nh \approx 1.6 \times
10^{22}$~cm$^{-2}$,][] {sun04}. The fitted column density for the
power-law model, $\nh = 3.4^{+0.6}_{-0.5} \times 10^{22}$~cm$^{-2}$,
is significantly larger than the integrated 21 cm Galactic value of
$2\times 10^{22}$~cm$^{-2}$ averaged over a $0.\arcdeg7 \times
0.\arcdeg7$ patch of the sky \citep{dic90}. The best fitted blackbody
model yielded a temperature of $kT_{\rm BB} = 0.44\pm0.03$~keV with a
fit statistic of $\chi^2_{\nu} = 0.9$ for 133 degrees of freedom (see
Fig.~\ref{specfig}).  The bolometric luminosity at $7.1$~kpc is
$L_{BB}({\rm bol}) = 3.7\times10^{33}$~ergs~s$^{-1}$, corresponding to
a blackbody area of $\approx 1.0 \times 10^{11}\,d_{7.1}^2$~cm$^2$ or
$\approx 0.5\%$ of the NS surface (see Table~\ref{modeltable}). The
addition of a second component to either spectral model is
unconstrained. Evidently the flux has remained steady between both
\xmm\ observations and the \chandra\ one \citep{sew03} in 2001.


\begin{deluxetable}{lccc}
\tablecolumns{4}
\tablewidth{240pt}
\tablecaption{\xmm\ Spectral Results for \psr}
\tablehead{
\colhead{Parameter}  &  \colhead{2004 Oct 18} & \colhead{2004 Oct 23} & \colhead{Sum}
}
\startdata
\multispan4{\hfill \hbox{Blackbody~Model}\hfill \vspace{3pt}}\\
\tableline
\vspace{3pt} $N_{\rm H}$ ($10^{22}$ cm$^{-2}$)
                                       & $1.5^{+0.5}_{-0.4}$   & $1.4^{+0.4}_{-0.3}$    & $1.5^{+0.3}_{-0.3}$ \\
$kT$ (keV)			       & $0.43^{+0.04}_{-0.04}$& $0.44^{+0.04}_{-0.03}$ & $0.43^{+0.03}_{-0.03}$ \\
pn BB Area (cm$^2$)		       & $1.0\times10^{11}$    & $1.0\times10^{11}$     & $1.0\times10^{11}$  \\
pn BB Flux                             & $1.9\times10^{-13}$   & $2.1\times10^{-13}$	& $2.0\times10^{-13}$ \\
pn $L_{\rm BB}$(bol)\tablenotemark{a}  & $3.6\times10^{33}$    & $3.5\times10^{33}$ 	& $3.6\times10^{33}$  \\
MOS BB Area (cm$^2$)		       & $1.1\times10^{11}$    & $1.0\times10^{11}$     & $1.0\times10^{11}$  \\
MOS BB Flux                            & $2.1\times10^{-13}$   & $2.2\times10^{-13}$	& $2.1\times10^{-13}$ \\
MOS $L_{\rm BB}$(bol)\tablenotemark{a} & $4.0\times10^{33}$    & $3.7\times10^{33}$ 	& $3.8\times10^{33}$  \\
$\chi^2_{\nu}$(dof)                    & 0.9(101)              & 0.7(103)               & 0.9(132)           \\
\cutinhead{Power-law~Model}
$N_{\rm H}$ ($10^{22}$ cm$^{-2}$)      & $3.4^{+0.8}_{-0.7}$   & $3.4^{+0.8}_{-0.7}$    & $3.4^{+0.6}_{-0.5}$ \\
$\Gamma$ 			       & $5.2^{+0.8}_{-0.8}$   & $5.2^{+0.7}_{-0.6}$    & $5.3^{+0.5}_{-0.5}$ \\
pn  PL Flux                            & $2.0\times10^{-13}$   & $2.1\times10^{-13}$    & $2.0\times10^{-13}$ \\
pn  $L_{\rm PL}$                       & $1.5\times10^{33}$    & $1.6\times10^{33}$ 	& $1.7\times10^{33}$  \\
MOS PL Flux                            & $2.1\times10^{-13}$   & $2.2\times10^{-13}$	& $2.1\times10^{-13}$ \\
MOS $L_{\rm PL}$                       & $1.6\times10^{33}$    & $1.7\times10^{33}$ 	& $1.8\times10^{33}$  \\
$\chi^2_{\nu}$(dof)                    & 0.8(101)               & 0.7(103)               & 0.9(132)            \\
\enddata

\tablecomments{\footnotesize Parameters are derived from a linked fit
to the EPIC~pn and MOS spectra, with the normalizations
independent. Uncertainties are 95\% confidence intervals for two interesting
parameters.  All fluxes are absorbed and given in units of ergs
cm$^{-2}$ s$^{-1}$; fluxes are fitted and computed in the 1--5 keV
energy band. Luminosities are unabsorbed, in units of ergs s$^{-1}$,
and derived for a distance $d = 7.1$~kpc. }
\tablenotetext{a}{\footnotesize Bolometric blackbody luminosity.}
\label{spectable}
\end{deluxetable}


\begin{deluxetable}{lc}
\tablecolumns{2}
\tablewidth{240pt}
\tablecaption{Adopted Spectral Model for \psr}
\tablehead{
\colhead{Parameter}  & \colhead{Value}
}
\startdata
$N_{\rm H}$ (cm$^{-2}$)  & $(1.5 \pm 0.3) \times 10^{22}$  \\
$kT$ (keV)	         & $0.44 \pm 0.03$                 \\
Blackbody emitting area $A_{BB}$ (cm$^2$) & $(1.0 \pm 0.4) \times 10^{11}$  \\
Blackbody radius $R_{BB}$ (km)            & $0.9 \pm 0.2$  \\
Absorbed 1.0--5.0 keV flux (ergs~cm$^{-2}$~s$^{-1}$) & $(2.1 \pm 0.1) \times 10^{-13}$ \\
Bolometric luminosity $L_{\rm BB}$(bol) (ergs~s$^{-1}$) & $(3.7 \pm 0.9) \times 10^{33}$  \\
$\chi^2_{\nu}$(dof)                       & 0.9(133)     \\
\enddata
\tablecomments{\footnotesize All parameters 
including the normalization are derived from a linked
fit to EPIC pn and MOS spectra using summed data from both
observations. Uncertainties are 95\% confidence intervals for two interesting
parameters. Luminosity derived for a distance $d = 7.1$~kpc.}
\label{modeltable}
\end{deluxetable} 


Since the spectral fit to the blackbody model implies that only a
small fraction of the NS surface is detected, we also derived an
independent upper limit to the effective blackbody temperature of the
entire NS surface.  Assuming $d=7.1$~kpc, $N_{\rm H} = 1.5 \times
10^{22}$~cm$^{-2}$, and a radius at infinity of 12~km, we compared
simulated spectra of increasing blackbody temperature with the data
until the predicted spectrum exceeded the observed flux in the lowest
energy bins by $3\sigma$.  The resulting upper limit is $\log
T^{\infty}_e < 6.24$, which, because of the large distance and column
density, is not constraining of NS cooling theory.

Although the blackbody is the preferred spectral model, we note that it
is not entirely consistent with the large pulsed fraction of the light
curve assumed to be coming from a small spot on the surface of the NS.
Light bending caused by the strong gravitational field of the
NS generally prevents the observation of such large modulation 
as is seen from \psr\ if the angular dependence of
the emitted intensity is isotropic, especially if there are
assumed to be two antipodal hot spots.  Instead, it is necessary
to include beaming due to anisotropic opacity in a strong magnetic
field \citep*{zav95,oze01} in order to reproduce such high pulsed fractions.
Such modeling is beyond the scope of this paper.  But if
the actual pulse shape of \psr\ turns out to be variable, as hinted
in Figure~\ref{timeplot}, then it may be necessary to consider an entirely
different mechanism, such as cyclotron emission from
low-luminosity accretion \citep{lan82}.  A broad cyclotron spectrum
emitted from above the NS surface may resemble a blackbody in shape
while allowing a higher degree of beaming.


\section{Optical Observation}

\begin{figure}[h]
\centerline{
\hfill
\psfig{figure=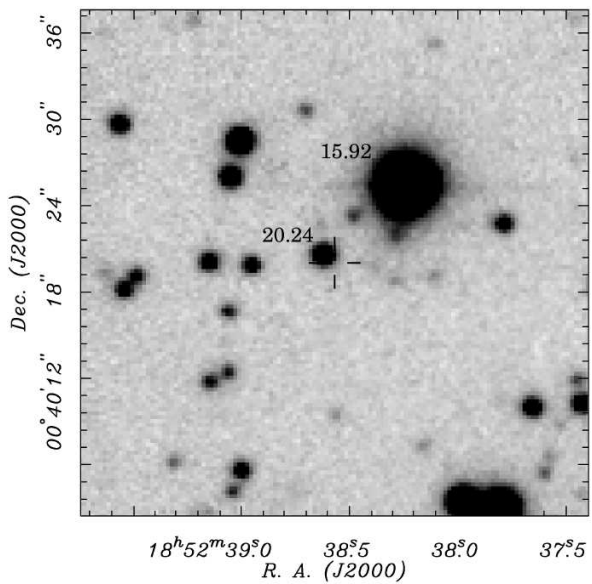,width=0.9\linewidth,angle=0}
\hfill
}
\caption{$R$-band CCD image at the location of PSR J1852+0040 in
Kes~79 obtained with the 2.4m Hiltner telescope.  Seeing is
$0.\!^{\prime\prime}88$.  The tic marks indicate the \chandra\
position of PSR J1852+0040, $18^{\rm h}52^{\rm m}38.\!^{\rm s}57,
+00^{\circ}40^{\prime}19.\!^{\prime\prime}8$ (J2000.0).  The magnitude
$R=15.92$ of the bright star is from the USNO-B1.0 catalog, while the
magnitude $R=20.24$ of the fainter star $0.\!^{\prime\prime}83$ from
the pulsar is scaled from the bright star.  The limiting ($3\sigma$)
magnitude of this image is $R=24.9$.
\label{optical_image}}
\end{figure}

Don Terndrup kindly obtained several CCD images of the field of \psr\
using the 2.4~m Hiltner telescope of the MDM Observatory on 2003 July
4 and 5.  A thinned, back-illuminated SITe $2048\times2048$ CCD with a
spatial scale of $0.\!^{\prime\prime}275$ per $24\,\mu$m pixel was
used with an $R$-band filter.  In Figure~\ref{optical_image}, we show
a combined 60 minute exposure that has seeing of
$0.\!^{\prime\prime}88$.  An astrometric solution for this image was
derived in the reference frame of the USNO-A2.0 catalog \citep{mon98}
using 65 stars that have an rms dispersion of $0.\!^{\prime\prime}49$.
The {\it Chandra} location of \psr\ is found to lie
$0.\!^{\prime\prime}83$ from a star of $R=20.24$ mag, the latter
estimated from the USNO-B1.0 calibration \citep{mon03}.  Given the
distance and extinction to \snr\ and the upper limit on the spin-down
power of the pulsar, we would not expect such a bright star to be its
optical counterpart.  Neither would we expect to detect the pulsar
even at the limiting magnitude of this image, which we estimate as
$R=24.9$.  The formal limiting magnitude at the location of the pulsar
is somewhat brighter than this because light from the nearby star
overlaps part of the $0.\!^{\prime\prime}6$ radius \chandra\ error
circle.


\section{Interpretation}

Using the limited information that we have about \psr\ we
can begin to evaluate possible models for its X-ray properties,
and the implications for the origin of CCOs in general.  In this
section we explore physical scenarios that are commonly
considered for those NSs that are not readily classified, and
conclude that none stands out as a preferred explanation of \psr.

\subsection{Rotation-Powered Pulsar}

Based on its period and spin-derived parameters,
the conventional interpretation of \psr\ is a rotation-powered
pulsar.  Its X-ray luminosity is less than
the inferred upper limit on its spin-down power, and it is
within the typical range for rotation-powered pulsars,
$10^{-4} < L_x/\dot E < 10^{-2}$ \citep{pos02}.
But if $L_x/\dot E$ of \psr\ is
found to be substantially larger than $10^{-2}$, that would be
unusual, and it would require a larger spin-down age, at odds with the
SNR association unless the pulsar were born with nearly its current
period.  The lower limit on the spin-down age is already greater than
the dynamical estimate of the SNR age (5.4--7.5~kyr, \citealt{sun04}).
If we adopt $\tau_{\rm SNR} < 7.5$~kyr as an upper limit to the true
age of the pulsar, then the dipole spin-down formula requires a birth
period $P_0 > 86$~ms and an initial spin-down power $\dot E_0 < 5
\times 10^{36}$ ergs~s$^{-1}$.

\nobreak \psr\ is clearly not among the energetic rotation-powered
pulsars, which we define as having $\dot E > E_{\rm c} \approx 4 \times
10^{36}$~erg~s$^{-1}$.  All pulsars with $\dot E > E_{\rm c}$ manifest
bright wind nebulae and have power-law-dominated spectra with photon
indices in the range $0.6 < \Gamma \simlt 2.1$ \citep{got04}.  Pulsars
just below $E_{\rm c}$ are found to have weak wind nebulae, if at all,
relative to the pulsar emission.  On average, their nebulae are
an order-of-magnitude less luminous than their super-critical
relatives, and their spectra are not yet well constrained.  There is no
indication of any hard extended emission around \psr, but deep
\chandra\ imaging is needed to exclude a weak nebula.  For less
energetic rotation-powered pulsars, the interpretation of their X-ray
spectra is not as clear because of the diversity of properties of
these objects.  Measurements of the brighter ones like PSR~J1709--4429
\citep*{got02,mcg04}, which is somewhat less luminous than \psr,
suggest a two-component spectrum composed of a blackbody model with
temperature $kT_{\rm BB} \approx 0.15$~keV in addition to harder
emission from a power-law model.  The temperature and luminosity 
of \psr\ are greater than those of middle-aged pulsars, which have
$L_x \sim 10^{31-32.5}$ ergs~s$^{-1}$ and $kT_{\rm BB} \leq 0.15$~keV,
The spectra of older pulsars tend to be dominated by even cooler blackbodies 
(see \citealt{del04}).  Thus, there is no single category of rotation-powered 
pulsar into which \psr\ fits neatly.

The emission of many intermediate-aged pulsars is dominated by
high-energy $\gamma$-rays, probably even those not yet detected
because of the limited sensitivity of EGRET.  The spin parameters of
\psr\ are not inconsistent with those of known $\gamma$-ray pulsars.
However, its location is confused with the EGRET source 3EG~J1856+0114
that is about $1^{\circ}$ from \snr, and is coincident with the
supernova remnant W44.  This EGRET source is hard but variable
\citep{nol03}.  We can assume that the flux of 3EG~J1856+0114,
$\approx 3.6\times 10^{-10}$~ergs~cm$^{-2}$~s$^{-1}$, is a conservative
upper limit on the gamma-ray flux of PSR J1852+0040.  Since the upper
limit on the spin-down flux $\dot E/4\pi d^2$ of PSR J1852+0040 is
$4\times 10^{-10}$~ergs~cm$^{-2}$~s$^{-1}$, it could be an as-yet
undetected $\gamma$-ray pulsar with an efficiency of a few percent,
typical of young or middle-aged pulsars.

Even though the X-ray luminosity of \psr\ is consistent with minimal
NS cooling curves for an age of $10^{3-4}$~yr \citep{pag04}, its
blackbody temperature implies an emitting area that is just $\approx
0.5\%$ of the NS surface. This is consistent with the highly
modulated pulse profile coming from a small rotating hot spot whose
measured temperature falls well above any reasonable NS cooling curve.
The most likely region for localized heating of the NS surface is at
the magnetic poles.  The canonical area for the polar cap is $A_{pc} =
{2\pi^2 R^3 / {P c}} \approx 1 \times 10^{10}$~cm$^2$. This is only
$10\%$ of the area implied by the fit to the X-ray spectrum using the
blackbody model. In the outer-gap model for $\gamma$-ray pulsars
\citep{wan98}, the X-ray luminosity of the hot polar cap is limited by
the Goldreich-Julian $e^{\pm}$ current flow of $\dot N_0 \approx 2
\times 10^{32}\,(P/0.105\,{\rm s})^{-2}\,(B/10^{12}\,{\rm
G})$~s$^{-1}$ depositing an average energy per particle of $E_f
\approx 4.3$~ergs.  The maximum luminosity is $L({\rm bol}) \approx f
E_f \dot N_0 < 4 \times 10^{32}$~ergs~s$^{-1}$.  Here the fraction $f$
of the current reaching the surface is set to 1/2, the maximum
possible estimated for a $\gamma$-ray pulsar near its death line.
While falling short of the observed X-ray luminosity of \psr\ by an
order-of-magnitude, this prediction applies only to a maximally
efficient $\gamma$-ray pulsar.  Either the $\gamma$-ray efficiency or
the $B_{\rm p}$ field is likely to be lower, so this mechanism is hard
pressed to account for the X-ray luminosity of \psr.  Polar-cap
heating models of \citet{har01,har02} predict even less X-ray
luminosity than \citet{wan98}. Taken at face value, all such
models fall short of predicting the apparent area, temperature, and
luminosity of the X-ray emission from \psr.

\subsection{Anomalous X-ray Pulsar}

While the temperature and luminosity of \psr\ are greater than those
of middle-aged pulsars, its luminosity is less
than those of AXPs, which have $L_x \sim 10^{34-35.5}$ ergs~s$^{-1}$
and thermal spectral components of $kT_{\rm BB} \simgt 0.4$~keV
\citep{mer02}.  The spectrum of \psr\ is suggestive of a magnetar of
low X-ray luminosity, perhaps like the quiescent state of the
transient AXP XTE~J1810--197 \citep{hal05}.  According to the magnetar
theory, the X-ray emission ultimately derives from the decay of an
enormous magnetic field ($B \simgt 4.4 \times 10^{13}$~G;
\citealt{dun96}).
Although \psr\ could be an ``anomalous,'' fast AXP, the implied magnetic
field strength is insufficient to power the observed X-ray luminosity
over the lifetime of the pulsar, estimated as $L_x \tau_{\rm SNR} \sim
8 \times 10^{44}$~ergs, since the available magnetic energy is only
$\approx B^2 R^3/6 = 3 \times 10^{43} (B/10^{13})^2$ ergs. More
detailed predictions invoking the magnetar theory (e.g., currents on
twisted magnetic field lines external to the star; \citealt*{tho02})
are similarly insufficient to sustain the observed X-ray emission.

\subsection{Accreting Binary}

Although binary NS X-ray transients in quiescence often have
luminosities similar to that of \psr, their spectra, as summarized,
e.g., by \citet{tom04}, are characterized as softer blackbodies
covering the full NS surface, rather than a small hot spot.  Even if
the hotter emission from \psr\ is hypothesized to come from residual
accretion, the current observations disfavor a binary scenario based
on its steady long-term flux, lack of orbital Doppler delay, and
absence of characteristic red noise in its timing spectrum.  The
unclassified star $<1^{\prime\prime}$ from the \chandra\ position,
while unlikely to be a binary companion of \psr, prevents us from
deriving a constraining upper limit on optical emission from either
the pulsar itself or a fall-back accretion disk.

\subsection{Fall-back Accretion}

Even if \psr\ possesses a fossil accretion disk, it may be unable to
accrete because the magnetospheric radius is $r_{\rm m} = 3 \times
10^8\,\mu_{30}^{4/7}\,(M/M_{\odot})^{1/7}\,
L_{37}^{-2/7}\,R_6^{-2/7}$~cm, which is $3 \times 10^9$~cm for an
assumed magnetic moment
$\mu = B_{\rm p}\,R^3/2 \approx 10^{30}$ G~cm$^3$ and an
observed $L = GM\dot m/R = 3.7 \times 10^{33}$ ergs~s$^{-1}$.
Therefore, the magnetic dipole pressure ejects any potential accreting
matter well outside the light cylinder radius, $r_{\ell} = cP/2\pi = 5
\times 10^8$~cm.  Only in the case of $B_{\rm p}$ as small as $7
\times 10^8$~G could \psr\ be a ``slow rotator,'' with $P \approx
P_{\rm eq}$, since the equilibrium (or minimum) period for disk
accretion is $P_{\rm eq} = 3.6\,\mu_{30}^{6/7}\,(M/M_{\odot})^{-2/7}\,
L_{37}^{-3/7}\,R_6^{-3/7}$~s.  While such a value of $B_{\rm p}$ is
common among low-mass X-ray binaries, it would be surprising for such
a young NS.

For an intermediate value of the magnetic field strength, \psr\ could be
in the propeller regime, $P < P_{\rm eq}$, in which matter is flung out from
the magnetospheric radius at a rate $\dot m$, which causes it to spin
down at a rate $\dot P \approx 2\,\dot m\,r_{\rm
m}^2\,I^{-1}\,P\,(1-P/P_{\rm eq})$ \citep[e.g.,][]{men99,zav04}, where $I
\approx 10^{45}$ g~cm$^2$ is the NS moment of inertia.  The observed
upper limit $\dot P < 7 \times 10^{-14}$ s~s$^{-1}$ sets an upper
limit of $\dot m < 3.7 \times 10^{16}\,(r_{\rm m}/10^8\,{\rm
cm})^{-2}$ g~s$^{-1}$ in the propeller scenario.  But in that case, it
is not clear how the highly pulsed, thermal X-ray emission is produced.
Even if the bulk of the fall-back material is ejected,
as little as $3 \times 10^{13}$ g~s$^{-1}$ accreting onto the NS surface
is sufficient to account for the observed luminosity.
This model was proposed by \cite{alp01} for CCOs in general, with the
additional requirement that their weak or undetected pulsations might
be washed out by an electron scattering corona accumulated from the much larger
$\dot m$ at the propeller radius.  Now, of course, this caveat no longer
applies in the case of \psr.  Indeed, the nearly 100\% modulation of its 
pulsed light curve in the first observation requires that the propeller 
mechanism create no significant X-ray emission or X-ray scattering
screen at the magnetospheric boundary.   As long as an accretion scenario
is being considered, there remains the associated possibility that the X-rays
are {\it not\/} surface thermal emission, but a broad cyclotron feature
that corresponds to a range of magnetic field strengths $B < 10^{12}$~G
just above the surface.

Because of the large distance and interstellar extinction to
\psr, our optical observation does not yet rule out the presence
of a fossil accretion disk.  We computed the optical emission of a
standard blackbody disk that is terminated at inner radius
$r_{\rm in} = r_{\rm m}$ and having values of $\dot m$ allowed
in the propeller scenario for \psr.  These important parameters
must satisfy the criterion 
$\dot m\,r_{\rm in}^2 < 3.7 \times 10^{32}$ g~s~$^{-1}$~cm$^2$
derived in the previous paragraph.  We include an extinction
of 6.8~mag in the $R$-band, corresponding to $N_{\rm H} =
1.5 \times 10^{22}$~cm$^{-2}$.  We find that face-on
disks with $r_{\rm in} > 5 \times 10^7$~cm,
an order of magnitude smaller than the light cylinder
radius, have $R > 25$~mag, which is fainter than our optical limit.
As long as there are significant
difficulties in understanding the temperature and (possibly variable)
modulation of the pulsed emission in the context of an isolated
NS, accretion from a fall-back disk remains a viable
option for \psr.

\subsection{Comparison with PSR J1210--5226}

Apparently, the Griffin-like properties of \psr\
are typical only of the CCOs, and are not easily accounted for in any
one theoretical framework.  Detailed comparison is possible only with
PSR J1210--5226, the 424~ms pulsar in PKS~1209--51/52 that has similar
luminosity but is modulated with a pulsed fraction of only $\approx
10\%$ \citep{zav00}. Although still classified as a CCO, the complex
spin-down behavior and spectrum of PSR J1210--5226 set it apart from
any other isolated NS. Whether it is unique or a typical CCO is not
yet clear. Its flux is found to be steady, but large variations of its
period derivative may imply a wide binary system ($P_{\rm orb} \sim
0.2-6$~yr), frequent large glitches,
or accretion from fall-back material \citep{zav04}.  In any
case, its wildly varying spin-down age is inconsistent with the
remnant age of $\sim 10$~kyr.  The X-ray spectrum shows broad
absorption features at $0.7$ and $1.4$~keV \citep{san02} and possibly
at $2.1$ and $2.8$~keV when fitted with a thermal blackbody continuum
model of $kT_{\rm BB} \approx 0.21$~keV and hard tail of $kT_{\rm BB}
\approx 0.40$~keV \citep{big03}. If \psr\ is like PSR~J1210--5226, we
would expect to see similar spectral features and timing behavior; the
limited data in hand are not strongly constraining.

\section{Conclusions and Future Work}

\psr\ is clearly a young NS associated with Kes~79, and apparently not
a close binary.  As a rotation-powered pulsar, the inferred upper
limit on its spin-down power is consistent with the absence of a
bright pulsar wind nebula.  From the age discrepancy between the
pulsar and SNR, it is possible that the initial spin-down power of
\psr\ is sub-critical ($\dot E_0 < \dot E_c$); this may prove to be a
defining property of the CCOs. What is clearly distinctive about \psr\
is its hot thermal emission and relatively high luminosity (for its
age and $\dot E$). These factors, along with the large pulse
modulation, suggest a rotating hot spot, but the area and luminosity
of this spot are greater by an order-of-magnitude than can be
explained by existing theories of isolated neutron stars.
Accretion of fall-back material from a fossil disk in the propeller
regime remains an option.   An alternative interpretation is a
low-luminosity AXP, but the inferred magnetic field and rotation
period are smaller than expected in this scenario. The X-rays from the
pulsar are not easily produced in the context of the magnetar theory
as the inferred $B$ field is insufficient to power the observed
emission over the implied age of the NS. These unexplained X-ray
properties of \psr\ may be symptomatic of the entire class of CCOs,
and may require intensive observational and theoretical efforts to
understand.  A radio detection would demonstrate cleanly
that \psr\ is a rotation-powered pulsar, but the absence of radio
pulsations would be inconclusive.  Frequent X-ray monitoring
will test for evidence of accretion torques and/or glitches, 
and establish whether the X-ray flux and pulse profile are variable.
By analogy with other isolated neutron stars that have
broad X-ray absorption features, a deeper spectral study of \psr\
may prove to be revealing.

\acknowledgments

This investigation is based on observations obtained with \xmm, an ESA
science mission with instruments and contributions directly funded by
ESA Member States and NASA. We thank Don Terndrup for obtaining the
optical image used in Figure~\ref{optical_image}. This work is
supported by NASA XMM grant NNG05GD46G.

\end{document}